# Raman spectroscopy study of pressure-induced phase transitions in single crystal CuInP$_2$S$_6$

Rahul Rao,[1]* Benjamin S. Conner,[2,3] Jie Jiang,[1,4] Ruth Pachter,[1] Michael A. Susner[1]

[1]Materials and Manufacturing Directorate, Air Force Research Laboratory, Wright-Patterson AFB, OH 45433 USA

[2]Sensors Directorate, Air Force Research Laboratory, Wright-Patterson AFB, OH 45433 USA

[3]National Research Council, Washington, D.C. 20001, USA

[4]UES Inc., Dayton OH 45432 USA

*Correspondence – rahul.rao.2@us.af.mil

## Abstract

Two-dimensional ferroic materials exhibit a variety of functional properties that can be tuned by temperature and pressure. CuInP$_2$S$_6$ is a layered material that is ferrielectric at room temperature and whose properties are a result of the unique structural arrangement of ordered Cu$^+$ and In$^{3+}$ cations within an [P$_2$S$_6$]$^{4-}$ anion backbone. Here, we investigate the effect of hydrostatic pressure on the structure of CIPS single crystals through a detailed Raman spectroscopy study. Analysis of the peak frequencies, intensities and widths reveals four high pressure regimes. At 5 GPa the material undergoes a monoclinic-trigonal phase transition. At higher pressures (5 – 12 GPa) we see Raman peak sharpening, possibly due to a change in the electronic structure, followed by an incommensurate phase between 12 and 17 GPa. Above 17 GPa we see evidence for metallization in the material. Our study shows that hydrostatic pressure could be used to tune the electronic and ferrielectric properties of Cu-deficient layered $CuInP_2S_6$.

$CuInP_2S_6$ (CIPS) is an emerging van der Waals two-dimensional (2D) material of interest due to its ferrielectric nature and above-room temperature Curie Temperature ($T_C \sim$ 315 K in bulk CIPS).[1,2] The ferrielectricity in CIPS derives from its structure (monoclinic crystal structure at room temperature, $Cc$ space group)[3,4] which consists of a framework of $[P_2S_6]^{4-}$ ethane-like ion groups wherein the S atoms form $S_6$ octahedra bridged by P-P dimers (1/3 of the octahedral sites) with hexagonally arranged Cu and In cations (remaining 2/3 of octahedral sites) within this framework (see schematic of a single layer in Fig. 1(a)). Below $T_C$, the Cu ions are located either in the upper or lower trigonal positions of the $S_6$ octahedra adjacent to the van der Waals gap while the In ions are located just below the center. This arrangement of the Cu ions leads to spontaneous electric polarization slightly offset by the counter movement of the In, thus making CIPS ferrielectric.[3] This preferred arrangement of metal cations is also responsible for the ferroic ordering observed in many related 2D metal thiophosphates (MTPs).[2,5,6]

Research on CIPS has accelerated significantly over the past few years with the observation of robust ferroelectricity in ultrathin flakes[7] and novel physics such as a quadruple well potential in Cu-deficient CIPS,[8] giant negative electrostriction [9], memristive capabilities[10] and large second-order non-linear optical effects.[11] Many of these applications are enabled by the high ionic mobility of the Cu ions, whose motion also results in the loss of ferroelectricity above the Curie temperature. It is therefore of great interest to investigate external means of tuning the structure and thereby the properties of CIPS. One method of achieving this is through pressure, specifically through hydrostatic compression. One of the basic questions that arises is whether pressure causes a loss of the 2D structure. In addition, since the Cu ions are known to be highly mobile, it is of interest to see how the application of hydrostatic pressure affects the mobility of these ions through the structural changes. Finally, it is of interest to study possible bandgap reduction and metallization at high pressures, similar to what has been recently observed in other 2D MTPs like $FePS_3$,[12] $CoPS_3$,[13] $CrPS_4$,[14] $SnP_2S_6$,[15] $Pb_2P_2S_6$,[16] and $CuInP_2S_6$-$In_{4/3}P_2S_6$ heterostructures.[17]

A handful of studies have performed high pressure measurements on CIPS and reported an increase in the Curie temperature on application of low pressures (up to ~370 K at 0.25 GPa),[18,19] the occurrence of a structural phase transition above 5 GPa,[20] as well as enhanced optoelectronic properties (photoresponse and second harmonic generation) up to 5 GPa.[21,22] However, it is still not entirely clear what happens to the CIPS structure at higher pressures (>5 GPa). Here we performed Raman spectroscopy under hydrostatic pressure (up to 20 GPa) on CIPS single crystals and uncovered at least two new structural transitions around 12 and 17 GPa. We attribute the two high-pressure transitions to isostructural changes in the electronic band structure, which in turn causes a decrease in the electronic bandgap and eventually the onset of metallicity in the material above 17 GPa.

The details of the crystal synthesis have been described elsewhere.[23] Briefly, $In_2S_3$ precursor (prepared from Alfa Aesar Puratronic elements, 99.999% purity, sealed in an evacuated fused silica ampoule and reacted at 950 °C for 48 h) was reacted with necessary quantities of Cu (Alfa Aesar Puratronic), P (Alfa Aesar Puratronic) and S (Alfa Aesar Puratronic) to obtain the correct stoichiometric composition of CIPS. The precursors were sealed in a fused silica ampoule with ~80 mg of $I_2$ and loaded into a tube furnace. The furnace was slowly ramped to 775 °C over a period of 24 hrs. and held at that temperature for 100 h. Following this, the sample was cooled at a rate of 20 °C/hr. The final composition was confirmed by electron dispersion spectroscopy (EDS) analysis using a Zeiss Gemini scanning electron microscope. Fig. 1(b) shows an optical image of a CIPS single crystal; our synthesis typically produces crystal sizes on the order of 1 mm.

High pressure Raman spectroscopy measurements were conducted using a Merrill-Bassett type diamond anvil cell (DAC) with Boehler-Almax anvils. The DAC was fitted with Type II-as diamonds with 500 µm culets, affording a maximum quasi-hydrostatic pressure of 20 GPa. A stainless steel gasket with a 90 µm aperture was placed between the anvils, and a CIPS crystal was loaded into the gap along with a small ruby crystal as a standard. The chamber was then filled with a 4:1 ratio methanol/ethanol solution which acted as the pressure transmitting medium.[24] Room temperature pressure-dependent Raman spectra were collected using a Renishaw inVia Raman microscope, with a 633 nm excitation laser focused through a 50 x magnification long-working distance objective lens on to the CIPS crystal inside the DAC. The spectra were collected with an excitation laser power of ~1 µW to minimize heating from the laser. The pressure was estimated using the R1 fluorescence peak of the ruby[25] and was exerted manually by driving three screws housed on the outside of the DAC. Spectral lineshape analysis was performed in Igor Pro by normalizing intensities with respect to the diamond Raman peak, followed a spline baseline subtraction and Lorentzian peak fitting to extract the peak frequencies and widths (full-width at half maximum intensity, FWHM).

The room temperature and ambient pressure Raman spectrum (collected with 633 nm excitation) from a CIPS single crystal is plotted in Fig. 1(c). Group theory analysis of the normal modes predicts $\Gamma_{vib.}$ = 28 A′ + 29 A″ modes,[26] where the non-degenerate A′ and A″ modes are both Raman and infrared active. Of these modes, 14 $A_g$ and 16 $B_g$ modes are expected to be Raman-active.[27] However, not all of these modes are observable in the room temperature ambient pressure Raman spectrum [Fig. 1(c)], possibly due to limited instrument resolution, depolarizing effects due to improper alignment of the single crystals, lattice defects etc. We are able to resolve 17 peaks in the Raman spectrum whose origins are as follows: peaks below 100 cm$^{-1}$ correspond to extended vibrational modes involving the Cu and In cations, peaks between 100 and 350 cm$^{-1}$ correspond to deformations of the octahedral cages and involve displacements of the S and P atoms. The sharp peak around 380 cm$^{-1}$ is dominated by out-of-plane P-P and S valence vibrations, and the high-frequency modes between 540-620 cm$^{-1}$ correspond mainly to out-of-plane P-P valence vibrations, with a minor contribution from the S atoms.[28] All 17 peaks along with their assignments are listed in supplementary material Table S1. The eigenvector displacements for four of the modes in Fig. 1(c) ($P_2$, $P_{10}$, $P_{13}$ and $P_{17}$) are shown in Fig. 1(d). These four modes are representative of the range of vibrational modes in CIPS, namely, metal ion vibrations ($P_2$), octahedral deformations ($P_8$), and P-S out-of-plane vibrations ($P_{13}$ and $P_{17}$). Their pressure-dependent frequencies, intensities and linewidths are discussed in greater detail below.

Raman spectra under compression are shown in the waterfall plot in Fig. 2. With the application of pressure, Raman peaks typically blueshift in frequency due to lattice anharmonicity effects.[29] As expected, all Raman peaks exhibit hardening up to the highest applied pressure. However, the biggest change in the spectra appears just under 5 GPa, where the number of peaks decrease drastically, especially between 100 – 350 cm$^{-1}$. At higher pressures, other than blueshifted peaks, the overall spectral shape remains unchanged up to the highest applied pressure (20 GPa), although we see several subtle changes in some of the peak widths and intensities. We performed our high pressure Raman study on three different CIPS crystals and found the same spectral trends in all three measurements. We did see that the Raman spectra recovered fully upon decompression to ambient pressure (supplementary material Fig. S1 shows the Raman spectrum after decompression, along with the initial spectrum at ambient pressure), indicating that the layered structure was preserved during pressurization, and that our observed structural changes at high pressure are reversible. Note that the Raman spectrum after decompression exhibits a slight broadening of peaks (e.g., 9.3 cm$^{-1}$ vs. 7 cm$^{-1}$ for $P_{13}$), indicating the presence of some residual stress after the release of pressure.

The changes peak frequency, intensity and width in the spectra above 5 GPa can be seen more clearly in Fig. 3, which shows a 2D heat map of spectral intensities as a function of pressure. Note that, in order to generate this plot, all spectra were normalized with respect to the intensity of the diamond Raman peak (~1333 cm$^{-1}$). Going up in pressure along the y-axis in Fig. 3, the structural phase transition around 5 GPa is clearly visible (indicated by the horizontal dashed white line in Fig. 3). Not only do the intensities of the low frequency peaks (between 100 – 200 cm-1) diminish severely, we also observe a significant broadening of the deformation mode ~210 cm$^{-1}$. This broadening appears to be a result of a few of those modes merging together above 5 GPa. Another prominent feature in the spectra is the redshift of the 265 cm$^{-1}$ deformation mode around 4.9 GPa, followed by a steady blueshift at higher pressures. Finally, the high frequency P-P stretching mode around 560 cm$^{-1}$ diminishes in intensity above a pressure of 5 GPa.

The dramatic changes in the Raman spectral features around 5 GPa have been observed previously and attributed to a monoclinic to trigonal ($R3$ space group) structural phase transition.[20] The trigonal phase has a higher symmetry compared to the monoclinic, hence the overall reduction in the number of Raman-active modes above the transition pressure. In addition to the transition above 5 GPa, we can also see slight jumps in the peak frequencies, especially $P_8$, $P_{10}$ and $P_{13}$ around 12 GPa. Interestingly, before the structural phase transition, we see a sharp decline in the intensities of the P-S deformation modes ~107 and 325 cm$^{-1}$ at 3.9 GPa, possibly due to loss of ferrielectricity at this pressure. This would imply a transition from the ferrielectric $Cc$ to paraelectric $C2/c$ structure above 3.9 GPa, followed by a further transition to the trigonal phase above 5 GPa. Further experimental and theoretical studies are needed to confirm this hypothesis. However, previous temperature-dependent Raman spectroscopy measurements did show that, upon quenching $CuInP_2S_6$-$In_{4/3}P_2S_6$ heterostructures from high temperatures, the deformation mode ~325 cm$^{-1}$ gets suppressed due to reordering of the grains and loss of ferrielectricity.[30]

At higher pressures, we see another slight shift in the peak widths and intensities around 17 GPa. These two higher pressure transitions are also indicated by the dashed lines in Fig. 3. To understand the spectral changes at higher pressures (12 and 17 GPa), we analyzed the frequencies, intensities and widths of several peaks. These are plotted in Fig. 4. Note that the errors from the peak fitting procedure have also been plotted in Fig. 4 and in some cases the errors bars are smaller than the data points. Starting with the lowest frequency peaks [$P_2$ and $P_4$, Fig. 4(a)], these correspond to extended vibrations mainly involving the Cu and In ions, as shown in the eigenvector schematic in Fig. 1(d). The frequencies of both peaks experience a jump to higher frequencies around the phase transition at 5 GPa (~6 and ~17 cm-1 for $P_2$ and $P_4$, respectively), and thereafter exhibit a steady blueshift up to the highest pressure. This trend is also seen for the frequencies of the other peaks [top panels in Figs. 4(b), 4(c) and 4(d)], although the frequencies of the P-S ($P_{10}$) and P-P ($P_{15}$, $P_{16}$) decrease slightly around a pressure of 12 GPa. We believe that this corresponds to an isostructural transition, and is clearer in the trends of the peak intensities and widths, discussed below.

The intensities of $P_2$, $P_4$ and $P_{16}$ decrease steadily [middle panels, Fig. 4(a)] and remain close to zero above 5 GPa, although these peaks do not completely disappear. A higher magnification view of the low frequency cation vibrational modes is shown in the supplementary material Fig. S2. The persistence of the low-frequency cation vibrational modes up to 20 GPa indicates that the arrangement of the metal cations did not change significantly under pressure. Perhaps the biggest change in intensity is observed for the high intensity P-S stretching mode ($P_{13}$). The intensity of this peak increases steadily between 5 – 12 GPa, followed by a sharp decline between 12 and 17 GPa, after which it increases again up to 20 GPa [middle panel, Fig. 4(c)]. The discontinuities in peak intensities are also observed for $P_8$, $P_{10}$, $P_{11}$ and $P_{15}$ especially around 17 GPa [middle panels, Figs. 4(b) and 4(d)], although they are not as sharp as the ones for $P_{13}$. Considering that $P_{10}$, $P_{13}$ and $P_{15}$ correspond mainly to out-of-plane stretching of the P and S atoms in the octahedra, the discontinuities in their intensities strongly suggests a direct consequence of decreasing layer spacing due to compression.

The discontinuities in peak intensities are also accompanied by concomitant changes in their peak widths. Similar to its intensity, the biggest change in the peak width occurs for $P_{13}$, which exhibits a steady decrease between 5 and 12 GPa, then a sharp increase of ~2 cm$^{-1}$ between 12 to 17 GPa, followed by a decrease back to ~7 cm$^{-1}$. This can also be seen for many of the other peaks ($P_2$, $P_4$, $P_8$ and $P_{10}$) although the trends are weaker. Peak broadening upon compression is expected due to lattice anharmonicity,[29] and has been observed during high pressure Raman spectroscopy studies of other 2D MTPs.[12] However, the anomalous sharpening of Raman peaks upon pressurization could be attributed to a couple of reasons: 1) Pressure-induced buckling of sulfur atoms due to a change in the stacking order along the *c* axis, stabilizing the phonon mode and resulting in a decrease in peak width, and/or 2) Changes in the band structure and a decrease in the bandgap. Indeed, a reduction of up to ca. 0.8 eV was demonstrated at high pressure.[22]

The former has been observed in a high pressure Raman spectroscopy study on FePS$_3$;[31] however, the buckling of S atoms preceded the structural phase transformation to the higher symmetry phase. In our case, CIPS already undergoes the monoclinic-trigonal phase transformation ~5 GPa, thus the decrease in peak widths between 5 and 12 GPa is likely due to the second reason, i.e., an electronic transition. Such an anomalous decrease in peak widths on pressurization has been observed previously and attributed to an electronic topological transition in materials like $Bi_2Se_3$, $Bi_2Te_3$ and $Sb_2Te_3$ and black phosphorus.[32–34] We have also observed a sharpening of peaks under pressure in $CuInP_2S_6$-$In_{4/3}P_2S_6$ heterostructures, and interestingly, the decrease occurred in the out-of-plane vibrational modes in the $CuInP_2S_6$-$In_{4/3}P_2S_6$ heterostructures.[17] This indicates that a similar change may occur in the band structure in $CuInP_2S_6$-$In_{4/3}P_2S_6$, albeit at a higher pressure (~13 GPa). In the case of CIPS, we attribute the simultaneous sharpening and increase in intensity of some of the out-of-plane vibrational modes ($P_8$, $P_{13}$) to a pressure-driven change in the band structure in the high symmetry trigonal phase. Interestingly, the highest frequency vibrational modes ~560 cm$^{-1}$ ($P_{15}$ and $P_{16}$) involve mainly out-of-plane vibrations of the P-P dimers, with a minor contribution from in-plane stretching of S atoms in the $S_6$ octahedra [see phonon eigenvector schematic in Fig. 1(d)]. $P_8$ and $P_{13}$, on the other hand, mainly involve in-plane stretching of S atoms with a minor contribution from out-of-plane vibrations of the P-P dimers [Fig. 1(d)]. Our observation of a greater effect of pressure on the linewidths of $P_8$ and $P_{13}$ compared to $P_{15}$ and $P_{16}$ therefore suggests that the change in the band structure upon compression is driven mainly by the weaker P-S bonds rather than the P-P bonds. Previous temperature-dependent Raman studies have also observed the P-S bonds to be affected by temperature much more than the P-P bonds.[3,35]

In a recent high pressure study on CIPS, Bu *et al.*[21] reported a decrease in electrical resistance above the monoclinic-trigonal phase transition (in their case it occurred ~3.9 GPa) pressure up to a pressure 7.6 GPa. At the same time, the bandgap of their CIPS sample decreased down to ~2.5 eV. Above 7.6 GPa, they observed an increase in electrical resistance up to ~15 GPa, followed by a sharp decline. The photocurrent increased dramatically as well above 15 GPa, suggestive of metallization of CIPS. Our results match Bu *et al.*'s results very well, although our transitions occur at slightly higher pressures. The sharp decline in the width of $P_{13}$ and increase in its intensity above 17 GPa therefore indicates metallization in our CIPS samples. Except for the highest frequency peak ($P_{16}$), all peaks exhibit a sharpening above 17 GPa, signifying a change in the electronic band structure of the entire material, and a possible conversion from a wide bandgap semiconductor to a metal at very high pressures (beyond our measurement range).

Our high pressure Raman spectroscopy study on single crystal CIPS revealed four regimes. The first is from ambient pressure up to ~5 GPa, above which the material undergoes a structural phase transition from the monoclinic to a higher symmetry trigonal phase. This transition is marked by a sudden decrease in the number of Raman peaks owing to the higher symmetry of the trigonal phase. Between 5 – 12 GPa, we see a sharpening of some of the Raman modes, especially those involving out-of-plane vibrations of the S and P atoms. The mode sharpening can be attributed to a change in the electronic band structure of CIPS due to a decrease in the bandgap. Between 12 – 17 GPa, these trends are reversed, with a reduction in peak intensities and broadening. We attribute this regime to an incommensurate structure, possibly caused by slight changes in the stacking between layers as well as the continuing decrease in layer spacing due to the compression. This incommensurate structure finally leads to the onset of metallization above 17 GPa, resulting in a marked decrease in the widths of the out-of-plane vibrational modes. Our study has revealed new features in the high pressure states of CIPS beyond the early reports of a single phase transition above 5 GPa,[20] and show that hydrostatic pressure may be used to reversibly tune the electronic structure as well as the ferrielectric properties of CIPS.

**Acknowledgements**

This research was funded by the Air Force Office of Scientific Research (AFOSR) grant LRIR 23RXCOR003 and AOARD-MOST Grant Number F4GGA21207H002. We also acknowledge support from the National Research Council's NRC Associateship program.

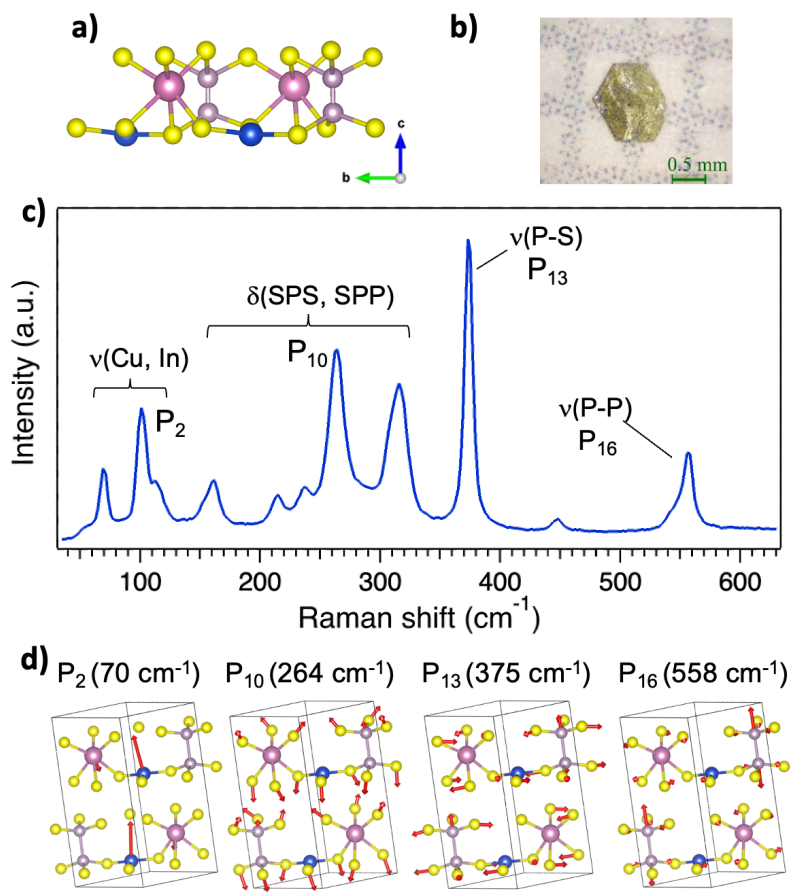

Fig. 1. a) Schematic of a single lamella of CIPS. The Cu, In, P and S atoms are in blue, pink, purple and yellow, respectively. b) Representative optical image of a CIPS single crystal used in this study. c) Room-temperature, ambient pressure Raman spectrum from CIPS obtained with 633 nm excitation. The types of vibrational modes are indicated in the figure. d) Phonon eigenvectors for four modes.

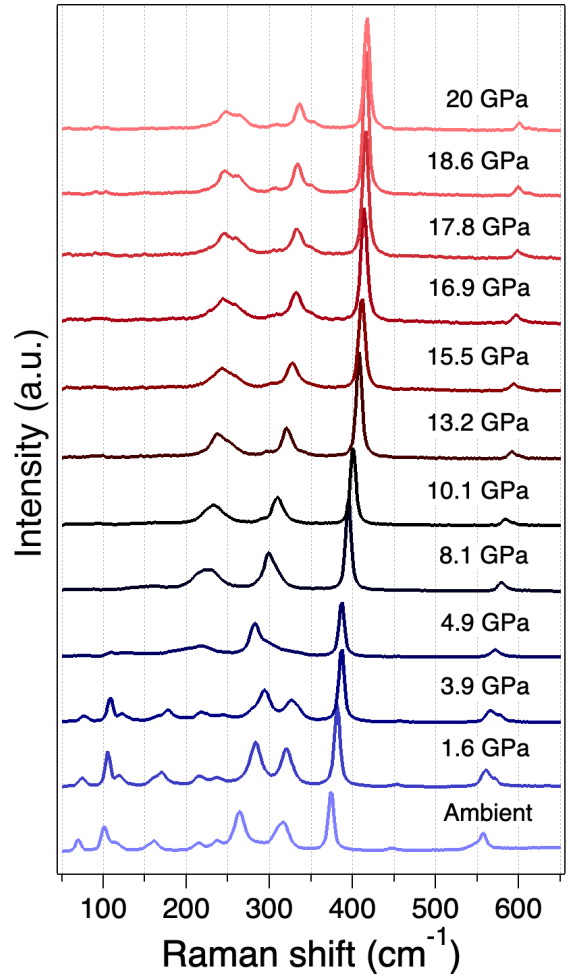

Figure 2. Waterfall plot showing Raman spectra upon compression (up to 20 GPa) from a CIPS single crystal.

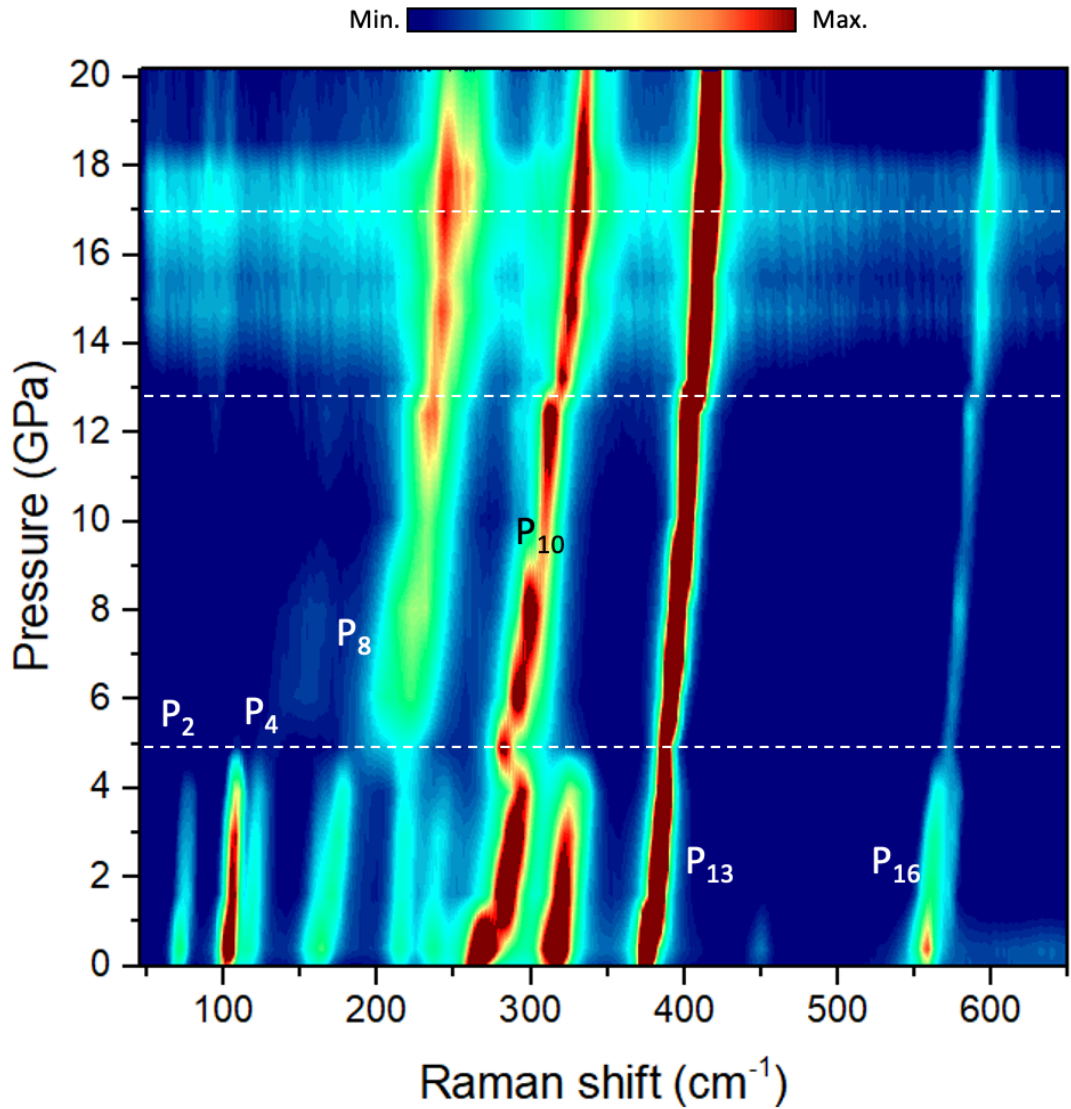

Fig. 3. 2D heat map showing the shift in Raman peak frequencies from CIPS upon compression. The three transitions around 5, 13 and 17 GPa are indicated as dashed horizontal lines. The color scale corresponds to peak intensities (counts) normalized to the diamond Raman peak.

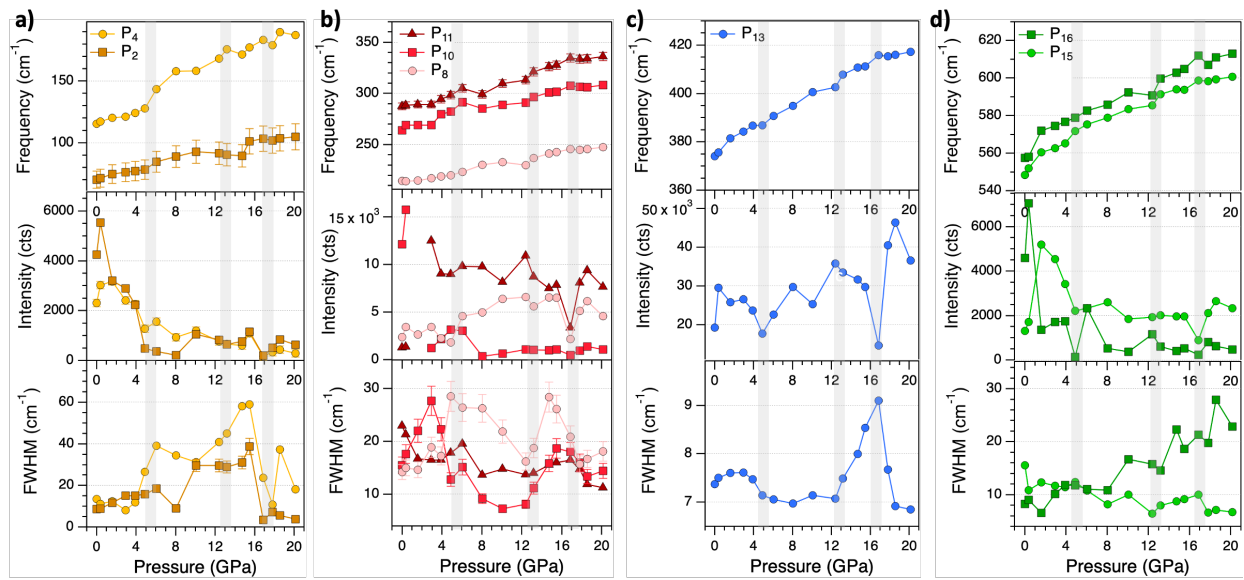

Fig. 4. a) – d) Pressure-dependent Raman peak frequencies (top row), intensities (middle row) and widths (FWHM, bottom row) for several peaks in the Raman spectrum of CIPS. The three phase transitions are shown by the shaded vertical boxes.